\documentclass[
 prb,
 amsmath,
 amssymb,
 reprint,
 superscriptaddress, 
 longbibliography,
 normalem
]{revtex4-2}

\usepackage{graphicx}
\usepackage{dcolumn}
\usepackage{bm}

\usepackage[utf8]{inputenc}
\usepackage[T1]{fontenc}
\usepackage{mathptmx}
\usepackage{etoolbox}

\usepackage{siunitx}
\usepackage{float}
\usepackage{xfrac}
\usepackage{hyperref}
\usepackage{xcolor}

\newcommand{\ket}[1]{|#1\rangle}
\newcommand{\bra}[1]{\langle#1|}
\newcommand{\sref}[2]{\hyperref[#1]{\ref*{#1}(#2)}}

\makeatletter
\def\@email#1#2{%
 \endgroup
 \patchcmd{\titleblock@produce}
  {\frontmatter@RRAPformat}
  {\frontmatter@RRAPformat{\produce@RRAP{*#1\href{mailto:#2}{#2}}}\frontmatter@RRAPformat}
  {}{}
}
\makeatother

\begin{document}

\title{Multiple Rabi rotations of trions in InGaAs quantum dots observed by 
photon echo spectroscopy with spatially shaped laser pulses}
\date{\today}

\author{S. Grisard}
\email{stefan.grisard@tu-dortmund.de}
\affiliation{Experimentelle Physik 2, Technische Universit\"at Dortmund, 44221 Dortmund, Germany}

\author{H. Rose}
\affiliation{Paderborn University, Department of Physics \& Institute for Photonic Quantum Systems (PhoQS), 33098 Paderborn, Germany}

\author{A.~V.~Trifonov}
\affiliation{Experimentelle Physik 2, Technische Universit\"at Dortmund, 44221 Dortmund, Germany}

\author{R. Reichhardt}
\affiliation{Experimentelle Physik 2, Technische Universit\"at Dortmund, 44221 Dortmund, Germany}

\author{D. E. Reiter}
\affiliation{Condensed Matter Theory, Technische Universit\"at Dortmund, 44221 Dortmund, Germany }

\author{M. Reichelt}
\affiliation{Paderborn University, Department of Physics \& Institute for Photonic Quantum Systems (PhoQS), 33098 Paderborn, Germany}

\author{C. Schneider}
\affiliation{Technische Physik, Universit\"at W\"urzburg, 97074 W\"urzburg, Germany}
\affiliation{Institute of Physics, University of Oldenburg, 26129 Oldenburg, Germany}

\author{M. Kamp}
\affiliation{Technische Physik, Universit\"at W\"urzburg, 97074 W\"urzburg, Germany}

\author{S. H\"ofling}
\affiliation{Technische Physik, Universit\"at W\"urzburg, 97074 W\"urzburg, Germany}

\author{M.~Bayer}
\affiliation{Experimentelle Physik 2, Technische Universit\"at Dortmund, 44221 Dortmund, Germany}

\author{T. Meier}
\affiliation{Paderborn University, Department of Physics \& Institute for Photonic Quantum Systems (PhoQS), 33098 Paderborn, Germany}
\author{I.~A.~Akimov}
\affiliation{Experimentelle Physik 2, Technische Universit\"at Dortmund, 44221 Dortmund, Germany}
\begin{abstract}
We study Rabi rotations arising in intensity-dependent photon echoes 
from an ensemble of self-assembled InGaAs quantum dots. 
To achieve a uniform distribution of 
intensities within the excited ensemble, we introduce flattop intensity profiles of  
picosecond laser pulses. 
This allows us to overcome the damping of Rabi 
rotations imposed by the spatial inhomogeneity of Rabi frequencies  
by a Gaussian laser profile. 
Using photon echo polarimetry, we distinguish between the coherent optical responses from exciton and trion ensembles.  
Here, we demonstrate that a photo-induced charging of the quantum dots leads to a significant  
reduction of the number of neutral quantum dots under resonant excitation with intensive  
optical pulses with areas exceeding $\sfrac{\pi}{2}$. 
The trion ensemble shows robust Rabi rotations when the
area of the refocussing pulse is increased up to 5.5$\pi$.
We analyze the remaining attenuation of Rabi rotations
by theoretical modeling of excitation induced dephasing,
inhomogeneity of dipole moments, and coupling to acoustic phonons.   
The latter is identified as the dominating mechanism resulting in 
a loss of optical coherence during the action of the involved optical pulses. 
\end{abstract}    

\maketitle

\section{Introduction}
Coherent control over inhomogeneously broadened  
ensembles of optically addressable qubits is one of the key challenges for the realization of  
high-capacity storage of quantum light states~\cite{lvovsky_optical_2009}.  
Quantum memory protocols in ensembles of qubits are typically based on the   
photon echo (PE) effect where quantum information stored in an  
ensemble is reemitted by optically inverting the phase evolution of 
individual emitters~\cite{tittel_photon-echo_2009}.
In this context, excitonic complexes in semiconductor
quantum dots (QDs) represent outstanding systems as they 
typically combine a large ratio between inhomogeneous and homogeneous linewidths 
with the possibility for sub-picosecond optical initialization, which 
makes them advantageous over atomic vapors~\cite{phillips_storage_2001, duan_long-distance_2001}. 
As recently demonstrated, the relatively short excitation 
lifetimes of QDs, that 
limit potential storage times, can be extended from the picosecond to the nanosecond timescale 
by transfer between optical and spin coherence~\cite{langer_access_2014, kosarev_extending_2022}.  
However, the use of high optical powers for a complete inversion of the ensemble 
leads to an irreversible loss of microscopical coherence due to the interaction between 
the QDs and acoustic phonons~\cite{muljarov_dephasing_2004}. A widely used method to quantify intensity-dependent 
decoherence mechanisms is the analysis of Rabi rotations as a function of pulse area. 

So far, most of the studies on Rabi rotations were performed on single 
QDs~\cite{stievater_rabi_2001, wigger_rabi_2018, monniello_excitation-induced_2013, zrenner_coherent_2002} because 
inhomogeneities of the Rabi frequency often hamper the observation of multiple Rabi cycles. 
Those inhomogeneities may arise from a spread of dipole moments (resulting from different QD sizes, compositions, etc.) 
or the coexistence of different 
excitonic complexes like excitons, trions, or 
biexcitons within the ensemble~\cite{suzuki_detuning_2018, borri_rabi_2002}. 
Furthermore, coupling to phonons results in a strong dependence on the detuning, 
which has to be taken into account 
when an ensemble is considered~\cite{klasen_optical_2021, quilter_phonon-assisted_2015}. 
Finally, the spatial inhomogeneity 
of the laser intensity profile that covers the QD ensemble induces 
as major source of dephasing that strongly hinders the experimental observation 
of Rabi rotations~\cite{poltavtsev_damping_2017}. 

In this paper, we study Rabi rotations arising in intensity-dependent PEs  
from an ensemble of self-assembled InGaAs QDs. We manipulate the spatial profiles of  
picosecond laser pulses to achieve a uniform distribution of 
intensities within the excited ensemble. 
This allows us to overcome the damping of Rabi 
rotations imposed by the spatial inhomogeneity of Rabi frequencies  
by a Gaussian laser profile while  
effectively using the available laser power. 
Using PE polarimetry, we study exciton and trion ensembles independently.  
Here, we demonstrate that photo-induced charging of the QDs leads to significant  
reduction of the number of neutral QDs under resonant excitation with intensive  
optical pulses with areas exceeding $\sfrac{\pi}{2}$. 
In this case, most of the QDs in the ensemble are charged and the main contribution to  
the PE signal is represented by trions that show robust Rabi rotations as a function of the area of the
refocussing pulse.
We analyze the experimentally observed damping of Rabi rotations
by theoretical modeling of excitation induced dephasing,
inhomogeneity of dipole moments, and phonon-assisted 
transitions between the dressed states during pulse action.

\section{Sample and Method}
\begin{figure}
    \centering
    \includegraphics[scale = 1]{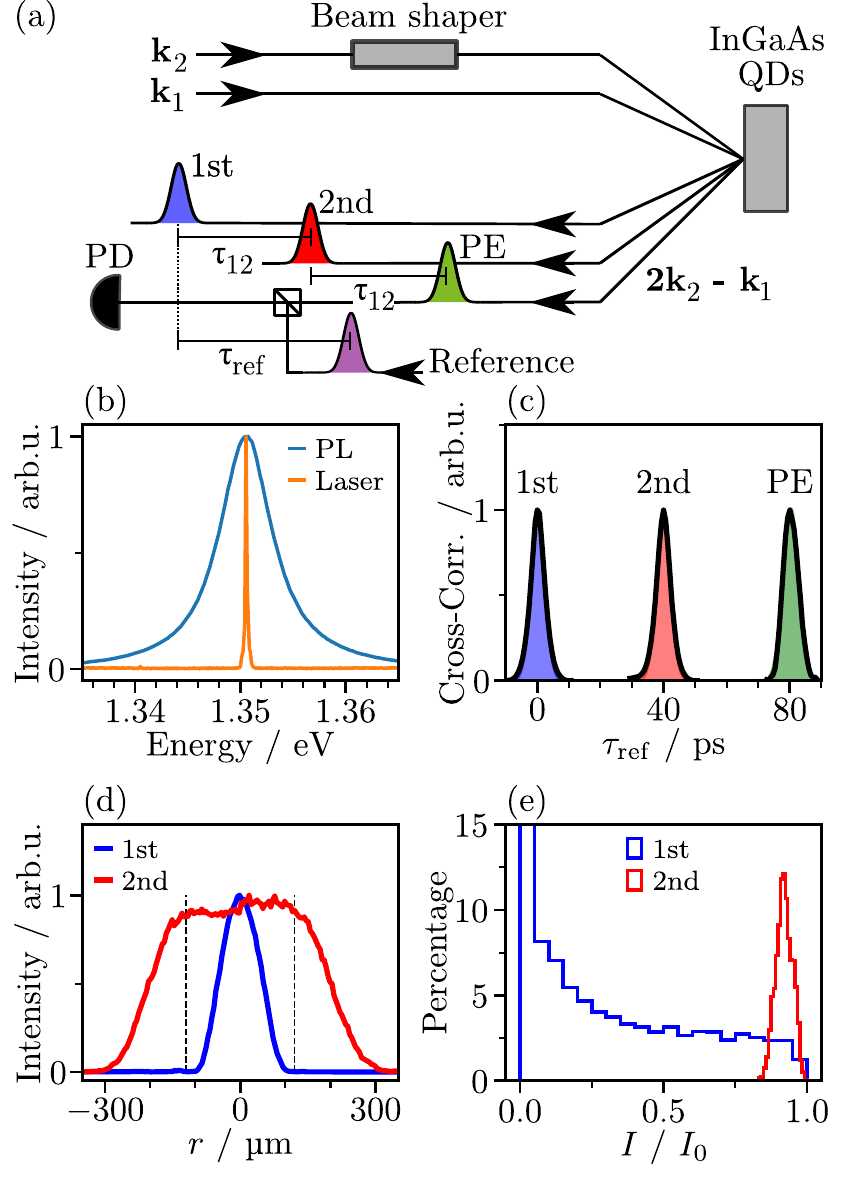}
    \caption{(a) Experimental scheme for detection of photon echo signals in reflection 
    geometry as described in the main text. (b) Laser spectrum and photoluminescence (PL) spectrum of the studied InGaAs QD sample placed in a AlGaAs microcavity. 
    The sample spectrum peaks at $\SI{1.351}{\eV}$ and has a FWHM of $\SI{5.9}{\milli\eV}$, 
    which results in a quality factor of 
    roughly 230. The \SI{0.3}{\milli\eV} 
    narrow spectrum of the laser is tuned to the central wavelength of the PL spectrum.
    (c) Experimental cross-correlations between the reference pulse and all involved pulses in the PE experiment.  
    The pulses have a duration of roughly \SI{3.8}{\pico\second}. 
    Since the laser spectrum selects a narrow sub-ensemble of QDs, 
    also the PE has a comparable duration to the other pulses. 
    (d) Cross-sections of the intensity profiles of first and second beam in the focal plane. 
    The profile of the second beam is manipulated to achieve a flattop 
    distribution in the focus while the first beam remains Gaussian. 
    In this way, we aim to overcome the damping of Rabi rotations by spatial inhomogeneity of Rabi frequencies. 
    (e) Distribution of intensities normalized to the maximum intensity $I_0$ 
    for the two beams within the radius shown by vertical dashed lines in (d).}
    \label{fig: figure01}
\end{figure}  

We study Rabi rotations
arising in intensity-dependent PEs on ensembles of
(In, Ga)As/GaAs QDs at a temperature of $\SI{2}{\kelvin}$.
The sample consists of a single layer of (In,Ga)As QDs with a density of 
\SI{1.8e9}{\centi\meter^{-2}}, surrounded by an AlGaAs $\lambda$-microcavity.
The photonic mode of the microcavity is in resonance with the QD ensemble 
at $\SI{1.351}{\eV}$ as we show in Figure~\sref{fig: figure01}{b} by the 
photoluminescence spectrum from the sample. The spectrum has a full width at half maximum (FWHM) of 
\SI{5.9}{\milli\eV}, which results in a quality factor of roughly 230. 
The same sample was studied before in 
\cite{kosarev_accurate_2020} and \cite{poltavtsev_photon_2016}, where further details can be found. A description of the sample fabrication can 
be found in~\cite{maier_bright_2014}. 

We schematically show the experimental PE setup in Fig.~\sref{fig: figure01}{a}. 
Here, two pulses with wavevectors $\mathbf{k}_1$ and $\mathbf{k}_2$ impinge on the sample temporally 
separated by an 
adjustable delay of $\tau_{12}$. 
The first pulse excites a macroscopic polarization that dephases quickly in 
inhomogeneously broadened systems such as QD ensembles. The second pulse inverts the 
phase evolution of each excitation independently, which leads to coherent emission 
in form of a PE at time $2\tau_{12}$ in the phase-matched direction $2\mathbf{k}_2 - \mathbf{k}_1$.
We experimentally detect the PE in reflection geometry using 
the optical heterodyne technique where we capture the interference 
between the weak signal of interest and a strong reference pulse on a photo diode (PD). 
By changing the delay $\tau_\text{ref}$ between first and reference pulse, 
we temporally resolve the photon-echo pulse.
Further details about the experimental technique 
can be found in reference~\mbox{\cite{poltavtsev_photon_2018}}. 
Figure~\sref{fig: figure01}{c} shows the measured temporal cross-correlations between the reference pulse and first, second, and PE pulses where 
we choose $\tau_{12} = \SI{40}{\pico\second}$. Note that 
we normalized the three independently measured peaks to their maximum.
The cross-correlations between the reference and first/second pulse both 
have a FWHM of $\SI{5.4}{\pico\second}$
reflecting the width of the pulse's amplitude profile $\SI{5.4}{\pico\second} / \sqrt{2} \approx \SI{3.8}{\pico\second}$. 
The temporal profile of the PE is given by the Fourier transform of the involved ensemble, hence allowing to extract the 
inhomogeneous decoherence time $T_2^*$. In our experiment however, 
the spectrum of the laser is more narrow than the cavity mode, as shown in Figure~\sref{fig: figure01}{b}. 
Thus, the PE duration $\SI{4.3}{\pico\second}$ is mainly determined by the pulse duration.
Next to the macroscopic decoherence time $T_2^*$, our method also allows to extract the 
homogeneous decoherence 
time $T_2$.
For that purpose, we capture the maximum value of the PE 
as a function of the delay $\tau_{12}$ by setting the reference time $\tau_\mathrm{ref} = 2 \tau_{12}$.
Results on measurements of $T_2$ are discussed below. 

Rabi oscillations are temporal rotations of the Bloch vector describing the coherent state of a two-level system
under action of an optical field.
If the optical field is in resonance with the two-level system, 
the frequency of the Rabi oscillations is given by $\Omega_R = \mu \mathcal{E} / \hbar$, 
with the dipole moment of the optical transition $\mu$, 
the light's electric field amplitude $\mathcal{E}$, and the reduced Planck's constant $\hbar$. 
For laser pulses, this phenomenon yields intensity-dependent Rabi 
rotations~\cite{stievater_rabi_2001, borri_rabi_2002}. 
In particular, those intensity-dependent Rabi rotations can be observed in a PE 
experiment as oscillations of the macroscopic polarization when the areas of the first or
second pulse are varied in a range strongly exceeding the 
$\chi^{(3)}$-regime of nonlinear optics~\cite{poltavtsev_photon_2016}. We refer to these oscillations 
simply as Rabi rotations throughout the article. 
The dimensionless pulse area~$A$ (or integrated Rabi frequency) is defined as
$A = \int \Omega_R(t) \mathrm{d}t$.
Upon changing the areas of first and second pulse (practically, by changing their intensity), 
the amplitude $P$ of the PE
oscillates according to~\cite{Berman2010}
\begin{equation}
    P \sim \sin\left(A_1\right) \sin^2\left(\frac{A_2}{2}\right).
    \label{eq: rabi_undamped} 
\end{equation}
Consequently, the PE runs through maxima for $A_1 = (2n + 1)\sfrac{\pi}{2}$, and $A_2 = (2n + 1)\pi$, 
respectively, where $n$ is an integer. The simple Equation~\eqref{eq: rabi_undamped} strictly holds only for 
delta-like pulses, 
while for finite pulses also the temporal shape of the PE can be 
modulated as a function of pulse area~\cite{poltavtsev_photon_2016}.  
For the present study, 
we restricted ourselves to Rabi rotations as a function of the second pulse's area $A_2$.
The area of the first pulse is fixed at $A_1 = \sfrac{\pi}{2}$,  
corresponding to the first maximum of the Rabi cycle. 
The choice of $A_2$-Rabi rotations is 
advantageous for the quantitative analysis of the damping of Rabi rotations
since sweeping the area of the first pulse also affects the temporal profile of the PE, as 
demonstrated in~\cite{poltavtsev_photon_2016}. 
Furthermore, because the sign of the PE amplitude is not affected by the area of the second pulse, 
$A_2$-rotations allow us to unambiguously distinguish 
a fading of Rabi rotations caused by an inhomogeneity of Rabi frequencies 
within the ensemble from a homogeneous drop of the microscopical polarization.

The distribution of intensities within the finite spot sizes 
of the laser pulses can introduce a source of fading of Rabi rotations~\cite{poltavtsev_damping_2017, kujiraoka_optical_2010}. 
Without modification, all involved laser profiles (first, second, reference) share a Gaussian distribution
with the same spot width $\sigma$. Consequently, the intensities vary within the spot diameter and the 
pulse areas in Equation~\eqref{eq: rabi_undamped} become dependent on the distance $r$ from the center of the spot. 
The signal detected by the heterodyne technique may be written in the following form 
\begin{equation}
    P(A_1, A_2) \sim \int_0^\infty r\, \underbrace{\mathrm{e}^{-\frac{r^2}{\sigma^2}}}_{\mathrm{Reference}}\sin\left[A_1(r)\right] \sin^2\left[\frac{A_2(r)}{2}\right]  \mathrm{d}r,
\end{equation}
where we assume radial symmetry of all spatial laser profiles.
When both, first and second beam have a Gaussian distribution, i.e. 
    $A_i = A_{i, 0}\,\mathrm{exp}\left(-\sfrac{r^2}{\sigma^2}\right)$
with a maximum pulse area $A_{i, 0}$, the Rabi rotations will be strongly damped since the spatial averaging 
is accompanied by a spread of pulse areas.

Here, we directly compare Rabi rotations with only Gaussian spots with the case 
that only the first pulse has a Gaussian distribution while the second beam is flat, 
i.e. $A_2(r) = A_{2, 0}$ within 
the area of spatial overlap between first and second. 
In this way, the Rabi rotations as a function of $A_{2, 0}$, 
while $A_{1, 0}$ is fixed, will not be damped by the effect of spatial averaging. 
To achieve this condition, we introduce a refractive beam shaper in the path 
of the second beam (Fig.~{\sref{fig: figure01}{a}}) that converts its Gaussian 
intensity profile to an Airy disk pattern.
By focusing the modified intensity distribution (mathematically, applying the Fourier transformation), 
we create a flattop intensity distribution as presented in Figure~\sref{fig: figure01}{d}.  
To quantify the difference between the two different intensity profiles, it is useful to consider
the distribution of intensities visualized by histograms in Figure~\sref{fig: figure01}{e}. 
For the Gaussian profile, the area within which we can find the same intensities grows with the distance from the center 
of the spot. Consequently, the intensities are spread over the full intensity range with an increasing trend 
towards zero. 
In contrast, the shaped profile narrowly concentrates all intensities around a mean value close 
to the maximum intensity $I_0$. 
Using the flat intensity profile for the second beam whereas the first beam 
remains Gaussian, we aim to significantly reduce the damping of Rabi rotations. 
A simpler approach to achieve 
the same degree of homogeneity is to use a significantly bigger Gaussian 
spot for the second beam~\cite{suzuki_detuning_2018}. 
However, a similar distribution 
of intensities as demonstrated in Figure~\sref{fig: figure01}{e}, 
is achieved by using a Gaussian beam with 
FWHM of roughly $\SI{1.2}{\milli\meter}$. 
In this case, however, only 4\% of the full intensity contributes to the signal.
Especially for the analysis of intensity-dependent damping mechanisms of Rabi rotations,
it is of major importance to efficiently use all optical power to reach high 
values of the pulse area.

\section{Excitons vs Trions}

\begin{figure}
    \centering
    \includegraphics[scale = 1]{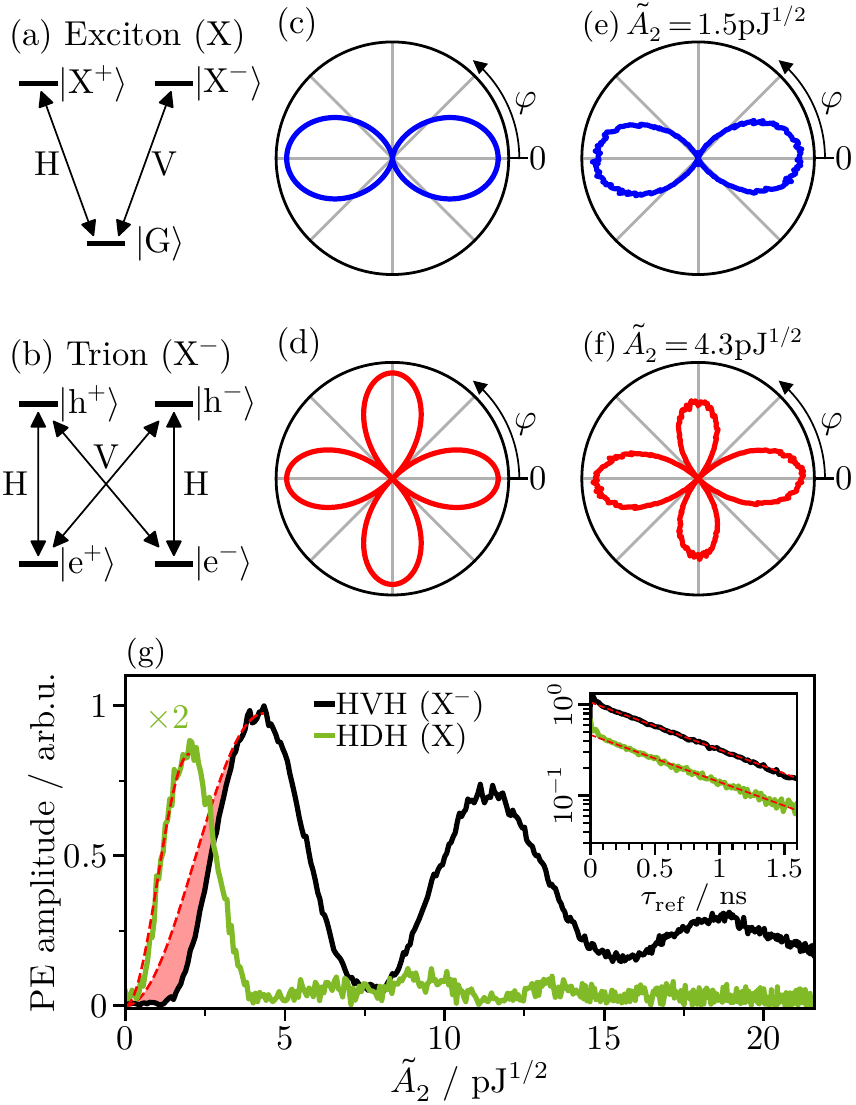}
    \caption{(a)~Schematic picture of the energy level arrangement of an exciton in linear polarization basis, where 
    $|\mathrm{G}\rangle$ is the ground state and 
    $|\mathrm{X}^{\pm}\rangle = \left(|\mathrm{X}^{\uparrow}\rangle \pm |\mathrm{X}^{\downarrow}\rangle\right) / \sqrt{2}$  
    are constructed from the spin up/down exciton states $|\mathrm{X}^{\uparrow/\downarrow}\rangle$.
    (b)~Trion energy level arrangement, where 
    $|\mathrm{e}^{\pm}\rangle = \left(|\mathrm{e}^{\uparrow}\rangle \pm |\mathrm{e}^{\downarrow}\rangle\right) / \sqrt{2}$ 
    denote the ground states with a resident electron and 
    $|\mathrm{h}^{\pm}\rangle = \left(|\mathrm{h}^{\uparrow}\rangle \pm |\mathrm{h}^{\downarrow}\rangle\right) / \sqrt{2}$  
    the excited trion states 
    constructed from the electron (hole) spin up/down states $|\mathrm{e}^{\uparrow/\downarrow}\rangle$ $\left(|\mathrm{h}^{\uparrow/\downarrow}\rangle\right)$. The labled arrows in (a) and (b) mark the allowed optical dipole transitions  
    in the linear polarization base $\left\{\mathrm{H},\,\mathrm{V}\right\}$. (c)/(d)~Modeled behavior of exciton and trion 
    within the PE polarimetry experiment, where the linear polarization $\varphi$ of the second 
    pulse is rotated with respect to the 
    horizontal polarization of the first and reference pulse. (e)/(f)~Measured polarization dependence 
    for two different values of the pulse area 
    of the second pulse as marked in the figure titles. (g)~PE amplitude 
    as a function of the pulse area $\tilde{A}_2$ (in units of square root energy per pulse), measured in the polarization configurations 
    HVH and HDH. Dashed lines mark the ideal $\sin^2(\sfrac{A_2}{2})$ dependence. Red area highlights the discrepancy 
    between the ideal 
    dependence (dashed line) and the measured data in the configuration HVH in the low power regime.
    Inset of (g) presents the PE decays measured in the configurations HVH (black) and HDH (green).}
    \label{fig: figure02} 
\end{figure} 
As a first step for the characterization of the QD ensemble, 
we study the origin of the exciton complexes giving rise to the PE signal. 
Different complexes such as neutral excitons (Figure~\sref{fig: figure02}{a}) or charged excitons (trions, Figure~\sref{fig: figure02}{b}) 
can possibly contribute to the signal. 
We exclude the contribution of biexcitons whose binding energy of roughly $\SI{3}{\milli\eV}$ is significantly 
larger than the spectral width of the ps-laser pulses~\cite{borri_rabi_2002}.
In the context of Rabi rotations, the coexistence of several exciton complexes may 
cause a source of inhomogeneity of dipole moments that effectively represents a source of damping.
However, the polarimetric PE technique allows to carefully analyze the contributions from different exciton complexes and, 
by appropriate choice of polarizations of the involved laser pulses, to independently address different sub-ensembles. 
In particular, we apply the PE polarimetry method as introduced in reference~\cite{poltavtsev_polarimetry_2019}. Here, 
the first pulse is horizontally polarized and the linear polarization angle $\varphi$ of the second pulse is rotated while 
the PE amplitude is detected in horizontal polarization. 
The choice of horizontal polarization for first and reference pulse is 
arbitrarily chosen since only the relative rotation between the involved polarizations 
is relevant. 
The PE amplitude arising from the exciton and trion 
exhibit a different functional dependence on the angle $\varphi$ due to the different 
energy level arrangement and optical transitions, visualized in Figures~\sref{fig: figure02}{a} and~\sref{fig: figure02}{b}. 
For the exciton, the PE amplitude is directly affected by the angle $\varphi$. 
For example, as follows from Figure~\sref{fig: figure02}{a}, 
the exciton does not give a response when first and second pulse are linearly cross-polarized ($\varphi = \sfrac{\pi}{2}$). 
The continuous dependence on the 
angle $\varphi$ thus reads as $\cos^2(\varphi)$.  
For the trion scheme in Figure~\sref{fig: figure02}{b}, however, a rotation of the second pulses polarization by the angle~$\varphi$ 
induces a rotation of the signal's polarization 
by the angle $2\varphi$, whereas the magnitude of the signal is unaffected. When the signal is detected in horizontal direction, 
this behavior leads to a $\left|\cos(2\varphi)\right|$ dependence. 
The dependences for exciton and trion 
are visualized by polar plots in Figures~\sref{fig: figure02}{c} and~\sref{fig: figure02}{d}. 
Note that two configurations 
play an essential role in distinguishing between excitons and trions. 
For $\varphi = \pi/2$, only the trion gives a signal, 
while for $\varphi = \pi/4$ only the exciton contributes. In the latter case,  
the PE signal is twice smaller due to polarization filtering in the detection.  
We denote the two configurations as HVH ($\varphi = \sfrac{\pi}{2}$) and HDH ($\varphi = \sfrac{\pi}{4}$) in the following,   
where the letters H, V, or D (horizontal, vertical, diagonal) mark the polarization of the first, second, and   
reference pulse, respectively.   

In the studied sample, we find that the polarimetric properties are strongly affected   
by the optical power of the involved pulses, as we demonstrate in Figures~\sref{fig: figure02}{e} and~\sref{fig: figure02}{f}.   
The power of the first pulse   
is fixed, while polar dependencies are measured for two values of the pulse area of the second pulse.
To measure the pulse area, we use the square root energy per pulse $\tilde{A}$. 
For $\tilde{A}_2 = \SI{1.5}{\pico\joule^{1 / 2}}$ the polar dependence resembles the two-leave rosette of a bare exciton (blue), 
while for $\tilde{A}_2 = \SI{4.3}{\pico\joule^{1 / 2}}$ we find a four-leave behavior like for the trion~(red). 
To study this observation in detail, we continuously measure the intensity dependence of 
the PE amplitude in the configurations 
HVH (trion) and HDH (exciton), presented in Figure~\sref{fig: figure02}{g}. 
In the HVH configuration, we find a pronounced Rabi rotations of the ensemble featuring 
three local maxima that we can associate 
with pulse areas of the second pulse $A_2 = \pi$, $3\pi$, $5\pi$ and two 
local minima for $A_2 = 2\pi$, $4\pi$, see Equation~\eqref{eq: rabi_undamped}. 
In contrast, the Rabi rotations captured in the configuration HDH appears strongly damped, only one maximum can 
be observed. 
Furthermore, the first maximum of the trion appears shifted towards higher optical powers by a factor 
of $\num{2.2}$ with respect to the exciton.  
A shift of the maxima could be caused by different dipole moments of exciton and trion. However, 
the dominant mechanisms for the different damping of the Rabi rotations of exciton and trions is revealed 
by considering the intensity dependence of the signal strength for the 
trion in the range up to the first maximum. 
For the exciton, the signal follows a $\sin^2(A_2 / 2)$ dependence as predicted by 
Equation~\eqref{eq: rabi_undamped}, highlighted by the red dashed line in Figure~\sref{fig: figure02}{g}. 
For the trion, however, we can observe a strong deviation that we highlight 
by the red area in Figure~\sref{fig: figure02}{g}. 
The signal in HVH rather follows a polynomial function of higher degree in the low power range 
indicating that another power-dependent contribution leads to an increase of the trion signal.
We associate this finding with a photo-charging of the QDs~\cite{dusanowski_optical_2022, johnsson_ultrafast_2019}. 
Increasing the 
optical power thus induces a rising of the number of singly charged QDs.
Accordingly, the number of neutral excitons is decreased, which explains 
the strong damping of the photon-echo signal measured in the configuration HDH. 
We note that the process of QD discharging acts on the timescale of 
microseconds and is thus significantly slower than the decoherence times of excitons and trions~\cite{kurzmann_auger_2016}.   
Therefore, it does not influence the temporal dynamics of PEs and their decay. 
However, it changes the ratio between the number of excitons and trions contributing to the signal.
If necessary, the same charging state of the QDs can be 
initialized in a controlled manner also by non-resonant illumination of the sample.

As a final remark to the polarimetry studies, we provide experimental values for 
the homogeneous linewidth associated with excitons and trions. 
In the inset of Figure~\sref{fig: figure02}{g}, 
we show the decays of the PE amplitude measured in HDH and HVH.
For both resonances, we extract $T_2 = \SI{0.83}{\nano\second}$ corresponding to 
a linewidth of $\Gamma = \sfrac{2\hbar}{T_2} = \SI{1.6}{\mu\eV}$. 
For the discussion of damping mechanisms of the Rabi rotations, we restrict ourselves 
to the polarization configuration HVH where we exclusively detect 
charged QDs. 
\section{Discussion of damping mechanisms}
\begin{figure}
    \centering
    \includegraphics[scale = 1]{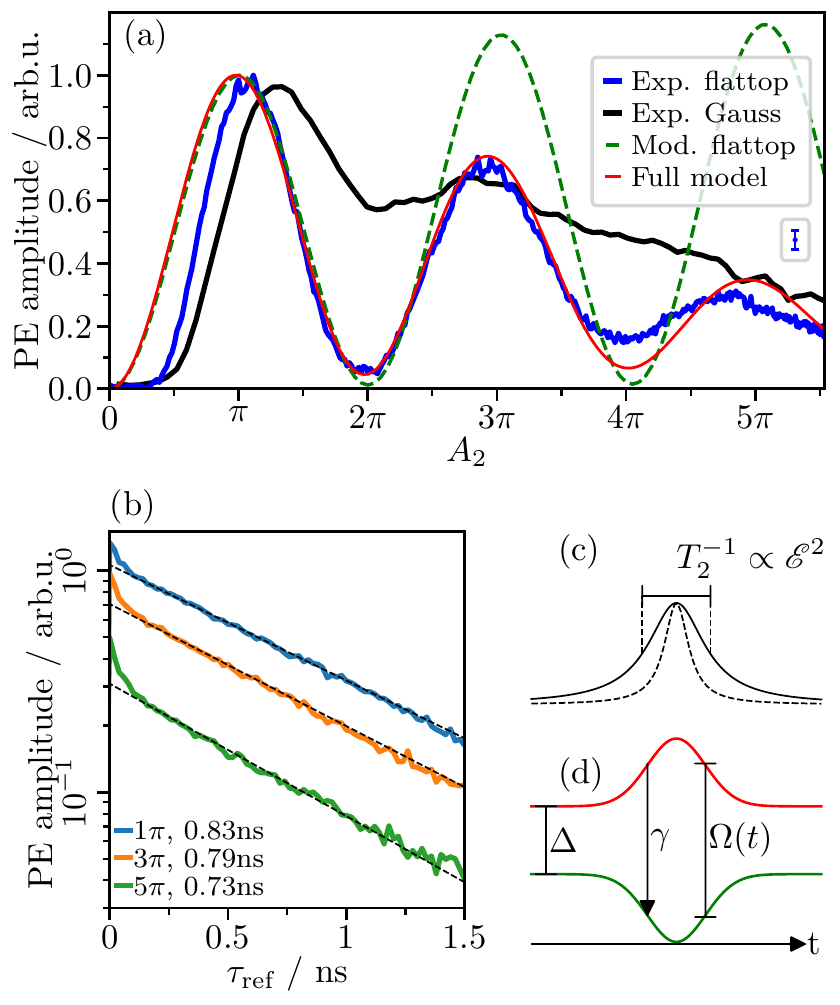}
    \caption{(a) Experimentally observed Rabi rotations as a function of the second pulse's area $A_2$ 
    for Gaussian and flattop intensity profiles. Red and green curve show results of our modeling procedure as described 
    in the text. 
    Note that the maximum uncertainty associated with the measured values of the PE amplitude 
    amounts to roughly $3\times 10^{-2}$, which we visualize by 
    the error bar shown below the legend in (a). 
    (b) Decays of the PE amplitude for $A_2 = \pi, \, 3\pi, \, 5\pi$ (local extrema of Rabi cycle), measured using 
    the flattop intensity profile for second beam. 
    Time values in the legend are results for the decoherence time $T_2$ extracted by fits (dashed lines).
    All experimental data shown in (a) and (b) where recorded in the polarization configuration HVH.
    (c) Schematic illustration 
    of excitation induced dephasing, that describes a broadening of the homogeneous linewidth, i.e. a shortening 
    of $T_2$ with increasing excitation power~$\sim\mathcal{E}^2$.
    (d)~Schematic illustration of the phonon model discussed in the main text. }
    \label{fig: figure03}
\end{figure} 
We aim to gain a deeper understanding into the underlying mechanisms that  
evoke the loss of macroscopic coherence in intensity-dependent PE experiments. 
First, we discuss the possible conclusions that can be drawn from a detailed 
consideration of the experimental data. 
Afterwards, we introduce a theoretical modeling procedure taking into account a 
variety of damping 
mechanisms that are either inherent to every single QD or are associated with the 
inhomogeneity of the ensemble. 
The comparison of model and experiment allows to quantify the contribution of 
each decoherence channel.
We want to stress that the usage of a flattop intensity profile 
is key to uncover internal sources of damping, which 
we demonstrate in Figure~\sref{fig: figure03}{a}. 
Here, we compare the Rabi rotations measured using a flattop intensity profile (blue)
with those measured with only Gaussian laser spots (black). 
Note that in the case of Gaussian spots, the Rabi cycle is 
strongly faded by the inhomogeneity of Rabi frequencies. 
The manipulation of the laser beam profile by means of a refractive beam profiler thus 
allows to strongly reduce the effect of spatial averaging. 
As we will show below, the effect of spatial inhomogeneity 
is negligible using the manipulated flattop profile as characterized in Figure~\sref{fig: figure01}{d}. 
We thus discuss further possible contributions to the damping mechanisms. 

The measured Rabi rotations in Figure~\sref{fig: figure03}{a} feature a strong drop 
of the contrast between the local maxima and minima with increasing pulse area. 
The contrast between the $\pi$ maximum and the $2\pi$ minimum is roughly \SI{90}{\percent}
whereas the contrast between the
$4\pi$ minimum and $5\pi$ maximum amounts to only $\SI{30}{\percent}$. 
This drop can be associated 
with an inhomogeneous broadening of dipole moments, which leads to a complete loss of contrast 
with increasing pulse area. Since the PE signal depends on $\sin^2(\sfrac{A_2}{2})$, 
the overall PE signal will reach a constant value of \SI{50}{\percent} of the maximum value. 
In our data, we can observe that the PE signal significantly drops below the value of 
\SI{50}{\percent} as the PE amplitude at $A_2 = 5\pi$ drops to roughly  
\SI{30}{\percent} of its maximum value. 
We therefore conclude that despite the damping mechanisms that are associated with the ensemble, 
we have to take into account decoherence effects that are inherent to a single~QD. 

A well known intrinsic contribution to the damping of Rabi rotations
is excitation induced dephasing (EID) that describes a broadening of the 
homogeneous spectral linewidth $\Gamma \sim T_2^{-1}$ with increasing excitation power, 
as we sketch in Figure~\sref{fig: figure03}{c}. 
Possible contributions to EID such as the coupling to phonons or 
wetting layer states
were intensively studied in 
several publications~\cite{borri_exciton_2005, luker_review_2019, ramsay_damping_2010, schneider_excitation-induced_2004}.   
Here, we only
aim to quantify the contribution of EID
by measuring the decoherence time $T_2$ as a function of excitation power.
In Figure~\sref{fig: figure03}{b} we measured the decay of the PE amplitude 
for $A_2 = \pi,\, 3\pi,\, 5\pi$. 
We notice that 
the decrease of the PE amplitude goes along with a gradual drop of $T_2$ 
from $\SI{0.83}{\nano\second}$ at $A_2 = \pi$ over $\SI{0.79}{\nano\second}$ at $A_2 = 3\pi$
to $\SI{0.73}{\nano\second}$ at $A_2 = 5\pi$.
We can conclude that EID is 
present in the studied sample and the accelerated decay directly damps the 
PE with a factor of $\exp\left[-\sfrac{\tau_\mathrm{ref}}{T_2(A_2)}\right]$, where 
$T_2(A_2)$ is the pulse-area-dependent decoherence time.
However, since we measure the Rabi rotations
for a relatively short delay of $\tau_\mathrm{ref} = \SI{80}{\pico\second}$, 
the contribution of EID on the drop of amplitude is negligible
as emerges from the 
estimation $e^{-\frac{80\mathrm{ps}}{T_2(5\pi)}} / e^{-\frac{80\mathrm{ps}}{T_2(\pi)}} \approx \SI{99}{\percent}$.
Instead, the strong decrease of the PE amplitude can solely be explained by
a drop of initial value of signal with increase of pulse area, which is in agreement with 
Figure~\sref{fig: figure03}{b}.
We therefore expand our consideration of decoherence  
beyond the broadening of the homogeneous linewidth and 
describe our full modeling procedure in the following.

We consider phonon-assisted relaxation processes during the action of the laser pulses 
that occur due to the optical dressing of the involved QD states. 
The importance of the effect has been demonstrated 
recently in phonon-assisted state preparation schemes of QDs~\cite{quilter_phonon-assisted_2015} or in the context of 
single photon-sources~\cite{thomas_bright_2021}. 
The idea of the process is illustrated in Figure~\sref{fig: figure03}{d} and will be described in the following.
The trion ensemble is considered 
as an ensemble of two-level systems (TLS) composed of the ground state $\ket{1}$ 
and the excited state $\ket{2}$, which is justified in the absence of a magnetic field. 
The density matrix $\rho$ of a single TLS is a $2\times 2$ matrix for a given basis.
We define $\rho_{12} = \bra{1}\rho\ket{2}$ as microscopic polarization 
and $\rho_{22} = \bra{2}\rho\ket{2}$ as occupation of the excited state. 
Computing the dynamic of the density matrix elements $\rho_{12}$ and $\rho_{22}$ 
in the bare-state (BS) basis leads to the optical Bloch equations (OBEs)~\cite{Allen.Eberly}.
The energy separation between the two states is given by the 
detuning $\Delta$ in rotating wave approximation.  
Upon interaction with a light field, the total Hamiltonian is diagonal in the dressed state (DS) basis with a 
modified energy splitting given by the generalized Rabi frequency $\Omega = \sqrt{\Omega^2_R + \Delta^2}$. 
A transition to the DS basis is advantageous when treating the interaction with longitudinal 
acoustic phonons, since their effect can be modeled approximately by introducing a loss rate $\gamma$ in the 
equations of motion~\cite{reiter_time-resolved_2017, klasen_optical_2021}. 
The dynamics of the density matrix elements in the DS basis 
is given by \cite{klasen_optical_2021}
\begin{align}
    \frac{\partial}{\partial t} \rho^{\mathrm{DS}}_{12} &= -i\Omega \rho^{\mathrm{DS}}_{12} -\frac{\gamma}{2}\rho^{\mathrm{DS}}_{12},\\
    \frac{\partial}{\partial t} \rho^{\mathrm{DS}}_{22} &= -\gamma \rho^{\mathrm{DS}}_{22}.
\end{align}
The phononic loss rate $\gamma$ is modeled as \cite{klasen_optical_2021}
\begin{align}
    \gamma &= \frac{\pi}{2}\Bigg(\frac{\Omega_R}{\Omega}\Bigg)^2 J(\Omega),\\
    J(\omega) &= A \omega^3 \exp\Bigg(-\frac{\omega^2}{\omega^2_c}\Bigg), \label{eq: phonon_density}
\end{align}
where $J(\Omega)$ is the phonon spectral density 
with the cut-off frequency $\omega_c$ and the amplitude $A$ of the coupling strength.
The phonon spectral density represents a measure for the 
efficiency of coupling between carriers and acoustical phonons for a given Rabi frequency $\Omega$.
Equation~\eqref{eq: phonon_density} 
assumes a spherical shape of the QDs, which may be a strong simplification for the shape of the studied QDs. However, 
by appropriate choice of the parameters $A$ and $\omega_c$, an arbitrary geometry 
of QDs may be sufficiently well approximated~\cite{luker_phonon_2017}. 
Therefore, we use $A$ and $\omega_c$ as fitting parameters of our modeling procedure.

Because in the considered approach the dephasing and relaxation induced by the phonons are only present during 
the excitation with an optical field, the free evolution of the system, temporally between the optical pulses, 
can be computed in the BS basis by 
OBEs in the rotating-wave approximation~\cite{Allen.Eberly}
\begin{align}
    \frac{\partial}{\partial t} \rho^{\mathrm{BS}}_{12} &= i \Delta \rho^{\mathrm{BS}}_{12},\\
    \frac{\partial}{\partial t} \rho^{\mathrm{BS}}_{22} &= 0.
\end{align}
The transformation between these bases is given by 
$\rho^{\mathrm{DS}} = V^{-1}\rho^{\mathrm{BS}} V$, 
where the matrix $V$ reads~\cite{klasen_optical_2021}
\begin{align}
    V = \frac{1}{\sqrt{2\Omega}}\begin{bmatrix}
\sqrt{\Omega - \Delta} & \sqrt{\Omega + \Delta}\\
\sqrt{\Omega + \Delta} & -\sqrt{\Omega - \Delta}
\end{bmatrix}.
\end{align}
Note that this transformation is defined for a fixed Rabi frequency $\Omega_R$. 
To describe pulses with Gaussian temporal shape and duration of $\SI{3.8}{\pico\second}$ 
that are used in the experiment, 
we approximate the time-dependent Rabi frequency $\Omega_R(t)$ 
by a step-like function with fixed values for finite temporal intervals.

To expand the described approach for an ensemble of TLS, we 
introduce three indices in the equations of motion 
\begin{align}
    \frac{\partial}{\partial t} [\rho^{\mathrm{DS}}_{12}]_\mathrm{i,d,s} &= -i\Omega_\mathrm{i,d,s} [\rho^{\mathrm{DS}}_{12}]_\mathrm{i,d,s} -\frac{\gamma_\mathrm{i, d, s}}{2}[\rho^{\mathrm{DS}}_{12}]_\mathrm{i,d,s}, \label{eq: extendedOBEs_0}\\
    \frac{\partial}{\partial t} [\rho^{\mathrm{DS}}_{22}]_\mathrm{i,d,s} &= -\gamma_\mathrm{i, d, s} [\rho^{\mathrm{DS}}_{22}]_\mathrm{i,d,s},\\
    \frac{\partial}{\partial t} [\rho^{\mathrm{BS}}_{12}]_\mathrm{i,d,s} &= i \Delta_\mathrm{i} [\rho^{\mathrm{BS}}_{12}]_\mathrm{i,d,s},\\
    \frac{\partial}{\partial t} [\rho^{\mathrm{BS}}_{22}]_\mathrm{i,d,s} &= 0.
    \label{eq: extendedOBEs}
\end{align}
In that way, we are able to account for inhomogeneous 
broadening of the QD ensemble (index i), randomly 
distributed dipole matrix elements (index d), and the 
spatial profile of the electric field (index s). 
The modeling procedure relies on numerical integration of Equations~\eqref{eq: extendedOBEs_0}-~\eqref{eq: extendedOBEs}. 
Subsequently, the ensemble averages over the three  
indices i, d, s are performed as described in the following.  
First, the macroscopic polarization is obtained by summation over the  
inhomogeneous broadening, given by a weight function $L$:  
\begin{align} 
    P_\mathrm{d,s}(t) = \mu_\mathrm{d} \sum_{\mathrm{i}} L(\Delta_j) p_\mathrm{i,d,s}(t) 
\end{align} 
For all calculations we assume a Lorentzian distribution of detunings 
$L(\Delta_\mathrm{i})$ with a FWHM of $\SI{5.9}{\milli\eV}$ 
as extracted from photoluminescence measurements, Figure~\sref{fig: figure01}{b}.  
Second, for the spread
of dipole moments (index d), we assume a Gaussian distribution $\tilde{G}(\mu_\mathrm{d})$ 
whose width is a free fitting parameter. The ensemble average is given by
\begin{equation}
    P_{s}(t) = \sum_d  \tilde{G}(\mu_\mathrm{d}) P_\mathrm{d,s}.
\end{equation}
Finally, the summation over the spatial profiles (index s) 
is carried out by taking into account the spatial profile of the reference beam $E^\mathrm{ref}_\mathrm{s}$, 
which equals the Gaussian spatial profile of the first pulse 
\begin{equation}
    P(t) = 2 \pi \sum_s r_s \Delta r E^\mathrm{ref}_\mathrm{s} P_\mathrm{s}(t),
\end{equation}
where we assumed radial symmetry. $\Delta r$ is the stepwidth for the discretization of the radial 
distance from the center of the spots $r$.
The final signal is obtained by convoluting $P(t)$
with the reference pulse, which is assumed to have the same
properties as the first pulse. 

The differential equations are solved with the fourth-order Runge-Kutta method  
with a step-width of $\mathrm{d}\tau = \SI{0.02}{\pico\second}$.   
The pulses are modeled as Gaussian function with a FWHM of \SI{3.8}{\pico\second}, 
discretized into 501 steps, which are sampled within an interval of $\SI{20}{\pico\second}$ centered around the pulse. 
For the inhomogeneous broadening, $N = 800$ TLS were taken into account covering 
the detuning range from $\SI{-15}{\milli\eV}$ to $\SI{+15}{\milli\eV}$.
The decoherence time $T_2$ is included a posteriori to the simulation of the sections 
at $\tau_\mathrm{ref} = 2\tau_{12}$, by multiplying with 
$\exp\left(-\frac{2\tau_{12}}{T_2(A_2)}\right)$, where $T_2(A_2)$ is a function that is obtained from a linear fit 
of the experimentally measured $T_2$ values, Figure~\sref{fig: figure03}{b}.

The described modeling procedure allows us to study individual decoherence mechanisms and to quantify 
their contribution to the overall damping. First, we use our model to 
specify the spatial homogeneity of the flattop intensity profile.
Figure~\sref{fig: figure03}{a} 
shows the modeled Rabi rotations where solely the experimental intensity distribution of the 
flattop laser profile 
is taken into account (green dashed line). 
No damping is observable within 
the measured intensity range. 
Therefore, we conclude
that our experimental method allows to overcome the effect of spatial averaging. 
Note that the modeled PE has a lower amplitude for $A_2 = \pi$ than for 
$A_2 = 3\pi$. This deviation from the simplified $\sin^2\left(\sfrac{A_2}{2}\right)$ dependence (Equation~\eqref{eq: rabi_undamped}) 
arises from the finite duration of the optical pulses as discussed above. 

Finally, we take into account all discussed damping mechanisms and 
optimize the free parameters of our model to fit the measured Rabi rotations. 
We find the best agreement 
between model and experiment for $A = \SI{0.012(1)}{\pico\second^2}$, $\omega_c = \SI{3.6(1)}{\tera\hertz}$, 
and a dipole inhomogeneity of $\SI{21(2)}{\percent}$ (FWHM), which are in reasonable agreement with 
results obtained from similar QD samples~\cite{quilter_phonon-assisted_2015, klasen_optical_2021, borri_rabi_2002}. 
The resulting Rabi rotation is plotted in Figure~\sref{fig: figure03}{a} together 
with the experimental data.  
Note that the modeling procedure does not take into account the photo charging effect that we discussed in the previous section. Therefore, the modeled curve deviates from the experimental data in the range $A_2 \lesssim \pi$.
Excellent agreement between model and experiment is found in the range $\pi \lesssim A_2 \lesssim 3.5\pi$, whereas the contrast of the measured oscillations is less pronounced for $A_2 \gtrsim 3.5\pi$ as described by our model. Nevertheless, the comparison between the experimental and modeled Rabi rotations in combination with the study of EID allows us to conclude that the coupling to acoustic phonons is the dominant mechanism for the loss of optical coherence inherent to a single dot.
For the used pulse durations of $\SI{3.8}{\pico\second}$, the largest pulse areas in our experiment $\sim 5\pi$ feature 
a mean Rabi frequency of roughly $\SI{4.1}{\tera\hertz}$, which is close to the maximum 
of the phonon spectral density~\eqref{eq: phonon_density} with the found parameters. Consequently, the 
loss of coherence acts very efficient in the regime of large pulse areas.
By use of longer ($\sim \SI{50}{\pico\second}$) or shorter ($\sim \SI{100}{\femto\second}$) 
pulses, the coupling to phonons could be strongly reduced~\cite{machnikowski_resonant_2004}.

\section{Conclusions}
We experimentally observed Rabi rotations in intensity-dependent PEs from an ensemble
of InGaAs QDs. Through application of the PE polarimetry method, we 
independently considered resonant excitation of 
excitons and trions and the resulting PE signals from these complexes. 
In the studied sample, the increase of optical power induces 
a rising number of charged QDs and 
decrease of neutral QDs due to photo-induced charging. 
Already for moderate excitation powers with area of $\pi$,
the majority of QDs are charged which allows to limit the consideration of 
one sub-ensemble where the coherent response stems from an ensemble of trions.
Rabi rotations on the trion population serve further as a tool to study intensity-dependent 
damping mechanisms that are associated with single dots and the ensemble. 
We successfully implemented a refractive beam shaper to achieve a uniform distribution of Rabi frequencies.
Thereby, we demonstrated Rabi rotations up to roughly 5.5$\pi$ allowing to 
discuss in detail different contributions to the damping. 
We found that the dominating source of damping inherent to a single dot is the efficient 
interaction with acoustic phonons.  
This interaction, however, is low for pulse areas up to $\pi$. 
Hence, more complex pulse arrangements including more than two pulses with areas up to $\pi$ as needed 
for quantum memory applications may be realized with high efficiency. 
Furthermore, the interaction with phonons for a given pulse area can be strongly 
reduced by extending or shortening 
the optical pulses.

\begin{acknowledgments}
We acknowledge financial support from the Deutsche Forschungsgemeinschaft (DFG) through the Collaborative 
Research Centre TRR 142 (project number 231447078, project A02) and 
from the German Ministry of Education and Research (BMBF) within the project
"QR.X" (FKZ: 16KISQ010).
We are grateful for the computation time grant provided by the PC$^2$ (Paderborn Center for Parallel Computing).
\end{acknowledgments}

\section*{Data Availability Statement}
The data that support the findings of this study are available from the corresponding author upon reasonable request.

\end{document}